\documentclass[11pt,a4paper]{article}
\usepackage{t1enc}
\usepackage[latin1]{inputenc}
\usepackage[english]{babel}
\usepackage{graphicx}

\begin{document}

\title{Nonlinear backreaction in a quantum mechanical SQUID}
\author{J.F.Ralph*$^{,1}$, T.D.Clark$^2$, M.J.Everitt$^2$, P.Stiffell$^2$  \\
$^1${\it Department of Electrical Engineering and Electronics, The University of Liverpool, } \\
{\it Brownlow Hill, Liverpool, L69 3GJ, United Kingdom. }\\ 
$^2${\it School of Engineering, The University of Sussex, } \\
{\it Falmer, Brighton, BN1 9QT, United Kingdom. }} 

\maketitle

\begin{abstract}
In this paper we discuss the coupling between a quantum mechanical superconducting
quantum interference device (SQUID) and an applied static magnetic 
field. We demonstrate that the backreaction of a SQUID on the
applied field can interfere with the ability to bias the SQUID at values of the
static (DC) magnetic flux at, or near to, transitions in the quantum mechanical SQUID.
\end{abstract}


*e-mail address: jfralph@liv.ac.uk

\section{Introduction}

There are a number of systems that are currently being considered as candidates for
the construction of qubits, quantum logic gates and quantum computers \cite{DiV97}. Some of the systems, 
notably atoms in magnetic traps \cite{Mon95} and nuclear magnetic resonance (NMR) systems \cite{Ger97}, have
had some success in performing the elementary operations that would be required in a large scale 
quantum computer. These systems benefit from the relatively weak coupling between the quantum
degrees of freedom used for the qubits and the external environment. This can result in coherence
times that are fairly long compared to the timescales used in the quantum calculations. However,
these systems are not necessarily seen as viable technologies for quantum computing in the
longer term. The difficulties involved in constructing large scale quantum circuits using such systems
are likely to be a limiting factor. A more realistic solution would be to develop qubit systems using 
solid state systems that would allow systems to be fabricated easily and repeatedly. The recent 
demonstration of macroscopic coherence in a SQUID ring \cite{Fri00,Wal00} (consisting of a thick 
superconducting ring containing one or more Josephson weak link devices) has added significant weight to the idea 
of using SQUIDs in quantum logic systems \cite{Bok97,Orl99,Chi00,Spi00,Sch00}, although other technologies are also 
being actively considered \cite{DiV97,Spi00}.

In this paper, we consider one aspect of the quantum mechanical SQUID that has previously been overlooked, and
we discuss how it may influence the construction and design of quantum logic gates based on SQUID devices.
The subject of this paper is the effect that the SQUID has on an applied magnetic field. Previous work
has concentrated on the appearance of nonlinear behavior in SQUID systems when they are coupled to 
radio-frequency oscillator circuits (`tank' circuits). This system has been investigated in both the classical
\cite{Soe85,Sch88} and the quantum regimes \cite{Ral92,Dig94}, and has been shown to contain a range
of interesting nonlinear effects. In the current work, we concentrate on a more fundamental problem: the nonlinear
effect of the SQUID on a static magnetic flux. In particular, we look at problems associated with fixing the classical
magnetic flux bias for a quantum mechanical SQUID at, or near, a quantum mechanical transition or resonance.

We present results that suggest that the backreaction of the SQUID on the static magnetic field can alter the
apparent shape of the quantum mechanical resonance even when the coupling between the field and the SQUID ring is 
weak. These results are important for quantum logic gates constructed using SQUIDs because
it is through a static (DC) magnetic flux that the behavior of a SQUID qubit is controlled 
\cite{Bok97,Orl99,Chi00,Spi00,Sch00}. There are some differences between the way in which the
flux bias is used, but the effect of the backreaction should remain where a quantum
resonance is being excited by an external time-varying field, although the size  of 
the effect will vary from system to system. This is particularly relevant for $\pi$-SQUIDs, where the adjacent 
wells in magentic flux are degenerate at zero applied flux \cite{Sch00}, which may reduce the significance
of the effect.

\section{Transitions in a quantum mechanical SQUID ring}

The behavior of a quantum mechanical SQUID ring in the presence of a time-dependent 
field is given by the time-dependent Schr\"{o}dinger equation (TDSE), which can be solved
using perturbative methods \cite{Fri00} or non-perturbative methods \cite{Cla98,Eve01}. 
In the later case, complex multi-photon transitions can be found for both semi-classical
\cite{Cla98} and fully quantum mechanical descriptions of the applied field \cite{Eve01}.
We adopt the non-perturbative, semi-classical approach described in \cite{Cla98}, although we
will restrict ourselves to single-photon, perturbative transitions for simplicity. However, the nonlinear 
analysis presented below is applicable to the non-perturbative, multi-photon transitions 
and to the transitions predicted using perturbative methods. In the case of the multi-photon
transitions, the complexity of the transitions would make it difficult to separate the nonlinear
effects from the transitions. Perturbative methods do provide an indication of the occurrence
of a transition and an estimate of the line-width of that transition, but they do not allow 
the shape of the resonance to be calculated, which is crucial for the determination of the nonlinear
backreaction.

A thick superconducting ring containing a single weak link ring (a radio-frequency (rf-)SQUID
ring) is often described in terms of a single macroscopic degree of freedom, $\Phi_{s}$, corresponding 
to the enclosed magnetic flux, with the electric displacement flux $Q_{s}$ playing the 
role of the conjugate momentum (strictly speaking the conjugate momentum is $-Q_{s}$, 
the commutator being given by $[Q_{s},\Phi_{s}]=i\hbar$). The Hamiltonian for 
the ring is given by,
\begin{equation}\label{ham}
H_{s}(\Phi_{x}(t))=\frac{Q_{s}^2}{2C_{s}}+\frac{(\Phi_{s}-\Phi_{x}(t))^2}{2\Lambda}-\hbar\nu\cos \left( \frac{2\pi\Phi_{s}}{\Phi_{0}}\right)
\end{equation}
where $C_{s}$ is the effective capacitance of the weak link, $\Lambda$ is the inductance of the ring,
$\nu$ is the (angular) tunneling frequency of the weak link (related to the critical current $I_{c}$ by
$\nu=I_{c}/2e$, where $2e$ is the charge of an electron pair), $\Phi_{0}=h/2e=2\times 10^{-15}$Wb 
is the magnetic flux quantum, and $\Phi_{x}(t)$ is the external magnetic flux applied 
to the ring. In this paper, we assume that the magnetic flux contains a time-dependent 
term to drive the resonance (typically at microwave frequencies $\Phi_{mw}(t)=\Phi_{mw}(t)\sin\left(\omega_{mw}t+\delta\right)$ 
where $\delta$ is an arbitrary phase), and a static DC magnetic flux to provide the bias point $\Phi_{dc}$. 
In experiments, such as those described in \cite{Fri00,Wal00,Whi98}, the microwaves are usually introduced via a coaxial cable 
acting as a transmission line, and the DC flux is applied by inductively coupling a current-carrying coil to the SQUID.

Figure 1 shows the first two (time-averaged) energy levels corresponding to the SQUID ground state and first excited state
for SQUID ring with inductance $\Lambda=3\times 10^{-10}$H, weak link capacitance $C_{s}=1\times 10^{-16}$F, 
$\hbar\nu=0.07 \Phi_{0}^2/\Lambda$ and subject to a microwave field of frequency $f_{mw}=\omega_{mw}/2\pi=144.7229$GHz
and amplitude $\Phi_{mw}=5\times 10^{-5}\Phi_{0}$, where the time-averaged energies are defined by,
\begin{equation}
\left<\left<E(\Phi_{dc})\right>\right>_{\kappa}=\frac{\omega_{mw}}{2\pi}\int^{2\pi/\omega_{mw}}_{0}dt\left<E(\Phi_{x}(t))\right>_{\kappa}
\end{equation}
and $\left<E(\Phi_{x}(t))\right>_{\kappa}$ is the instantaneous energy eigenvalue of the $\kappa$'th instantaneous
energy state of the Hamiltonian (\ref{ham}), as described in \cite{Cla98}, and correspond to the Floquet quasi-energies
that are used extensively in quantum optics \cite{Coh92}. For systems where the quantum transitions are at
microwave frequencies, it is assumed that it is the time-averaged energies (or the corresponding time-averaged
screening currents, see below) that induce a nonlinear backreaction in the DC coil. This approximation is based on the
assumption that the fluctuations in the DC magnetic flux occur at frequencies much lower than those present in the SQUID or the 
applied microwaves. 
\begin{figure}
\begin{center}
\includegraphics[height=6cm]{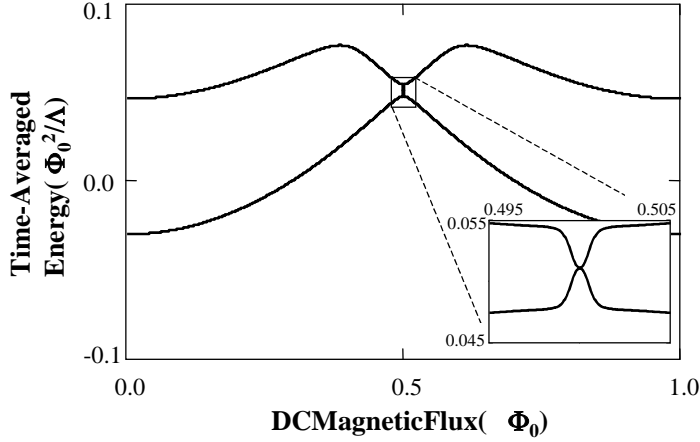}
\end{center}
\caption{Time-averaged energy levels for the first two energy states of an rf-SQUID ring, showing
a perturbative, single-photon transition at $\Phi_{dc}=0.5 \Phi_{0}$. The inset shows the same
transition in more detail. (The system parameters are given in the text).}
\end{figure}

\section{The effect of the backreaction on the DC bias}

The time-averaged energies shown in figure 1 are calculated assuming that the DC bias can be set at a desired
value with an arbitrary accuracy. At first glance one might assume that, in a real system, there will be some noise 
that will tend to `blur' out any very small features but, as long as the magnetic flux noise is small compared to 
the size of the features, the general form should be preserved. However, this is based on the assumption that it
is the static magnetic flux that is the control parameter. In practice, it is not the DC magnetic flux that is set, 
rather it is the DC current that flows through the coil that is fixed. The current flowing in the DC coil induces a flux
that couples to the SQUID, which induces a screening current in the SQUID ring that couples back to the DC coil
modifying the true value of the applied DC flux. This backreaction effect in quantum mechanical SQUID rings has 
been studied in the context of an rf-oscillator/SQUID system, where it is possible to derive
an equation of motion for a classical oscillator in the presence of an rf-SQUID that includes
the effect of coupling to all orders \cite{Ral92,Dig94}. We model the DC coil as an oscillator
(for the moment at least), and use the definition of the mutual inductance 
$M$ between a SQUID and an external inductive circuit,
\begin{equation}\label{fluxdef1}
\Phi_{s}=\Lambda I_{s}+M I_{t}
\end{equation}
\begin{equation}\label{fluxdef2}
\Phi_{t}=L_{t} I_{t}+M I_{s}
\end{equation}
where $\Phi_{t}$ is the magnetic flux in the oscillator,  
and the oscillator is characterized by a capacitance $C_{t}$, an inductance $L_{t}$. The Hamiltonian for the
combined system can be written in the form \cite{Ral92},
\begin{eqnarray}\label{ham2}
H&=&\frac{Q_{t}^2}{2C_{t}}+\frac{\Phi_{t}^2}{2L_{t}(1-K^2)}- \Phi_{t}I_{in}+\frac{K^2 \Phi_{t}\Phi_{s}}{M(1-K^2)} \nonumber \\
&&+\frac{Q_{s}^2}{2C_{s}}+\frac{\Phi_{s}^2}{2\Lambda(1-K^2)}-\hbar\nu\cos \left( \frac{2\pi\Phi_{s}}{\Phi_{0}}\right) \nonumber \\
&=&\frac{Q_{t}^2}{2C_{t}}+\frac{\Phi_{t}^2}{2L_{t}}- \Phi_{t}I_{in} \nonumber \\
&&+\frac{Q_{s}^2}{2C_{s}}+\frac{(\Phi_{s}-\mu\Phi_{t})^2}{2\Lambda(1-K^2)}-\hbar\nu\cos \left( \frac{2\pi\Phi_{s}}{\Phi_{0}}\right)\nonumber \\
&=&H_{t}(I_{in})+H_{s}(\mu\Phi_{t})
\end{eqnarray}
where $I_{in}$ is the external current applied to the oscillator and the coupling coefficients are given by $K^2=M^2/\Lambda L_t$
and $\mu=M/L_{t}$. From the Hamiltonian (\ref{ham2}), we can see that the effect of the coupling on the SQUID can be represented
as shifting the effective inductance of the SQUID ring by a factor $(1-K^2)$, $\Lambda\rightarrow\Lambda(1-K^2)$.

Averaging over the quantum behaviour, it is then possible to derive a classical equation of motion for the 
magnetic flux in the oscillator coil as a function of the applied current, 
\begin{equation}\label{osc}
C_t \frac{d^2\Phi_t}{dt^2}+\frac{1}{R_t} \frac{d\Phi_t}{dt}+\frac{\Phi_t}{L_t}=
I_{in}+\frac{\mu\left<\left<I_{S}(\mu\Phi_t)\right>\right>_{\kappa}}{(1-K^2)}
\end{equation}
where we have inserted a resistance $R_{t}$, and the time-averaged screening current in the SQUID ring is
calculated using the bare (unrenormalised) inductance of the SQUID ring $\Lambda$, leading to the
$1/(1-K^2)$ factor in the last term. (In reference \cite{Ral92} the average screening current in the
ring is calculated using the value of the renormalised inductance, which removes the multiplicative 
factor, but does not change the behaviour predicted by the equation). 
The inclusion of the time-averaged screening current,
\begin{equation}\label{sc}
\left<\left<I_{S}(\Phi_{dc})\right>\right>_{\kappa}=-\frac{\left<\left<\Phi_{s}(\Phi_{dc})\right>\right>_{\kappa}}{\Lambda}
=-\frac{\partial \left<\left<E(\Phi_{dc})\right>\right>_{\kappa}}{\partial \Phi_{dc}}
\end{equation}
is equivalent to the use of the Born-Oppenheimer approximation \cite{Ral92,Bor26} that is used in atomic and molecular calculations
for systems that vary over very different time scales (e.g. it is used to separate the slow dynamics of nuclei from the
very fast dynamics of electrons). In this situation, the Born-Oppenheimer approximation is used to separate the 
dynamics of the SQUID/microwave system from the dynamics associated with the DC coil. 

Since the subject of this paper is the behavior of the DC magnetic flux coil and all fluctuations associated
with the DC coil are assumed to have a very low frequency, we can approximate the oscillator equation (\ref{osc}) by,
\begin{equation}\label{dc}
\frac{\Phi_{dc}}{L_{dc}}=I_{dc}+\frac{\mu\left<\left<I_{S}(\mu\Phi_{dc})\right>\right>_{\kappa}}{(1-K^2)}
\end{equation}
where the quantities now relate to the DC coil and the DC current that is applied to it. This is the equation that
we can use to determine how much static flux couples to the SQUID ring. It is nonlinear, so that the DC flux 
is not necessarily proportional to the applied current. For a fixed current level we can solve this equation to
find the DC flux level that couples to the SQUID. Given a range of these DC values, it is possible to use
the calculated, time-averaged energy levels to predict the apparent energy level structure. 
Figure 2 shows the time-averaged energy levels from the inset from figure 1 as a function of the applied DC current 
for several different coupling strengths, including the original energies for comparison.
\begin{figure}
\begin{center}
\includegraphics[height=16cm]{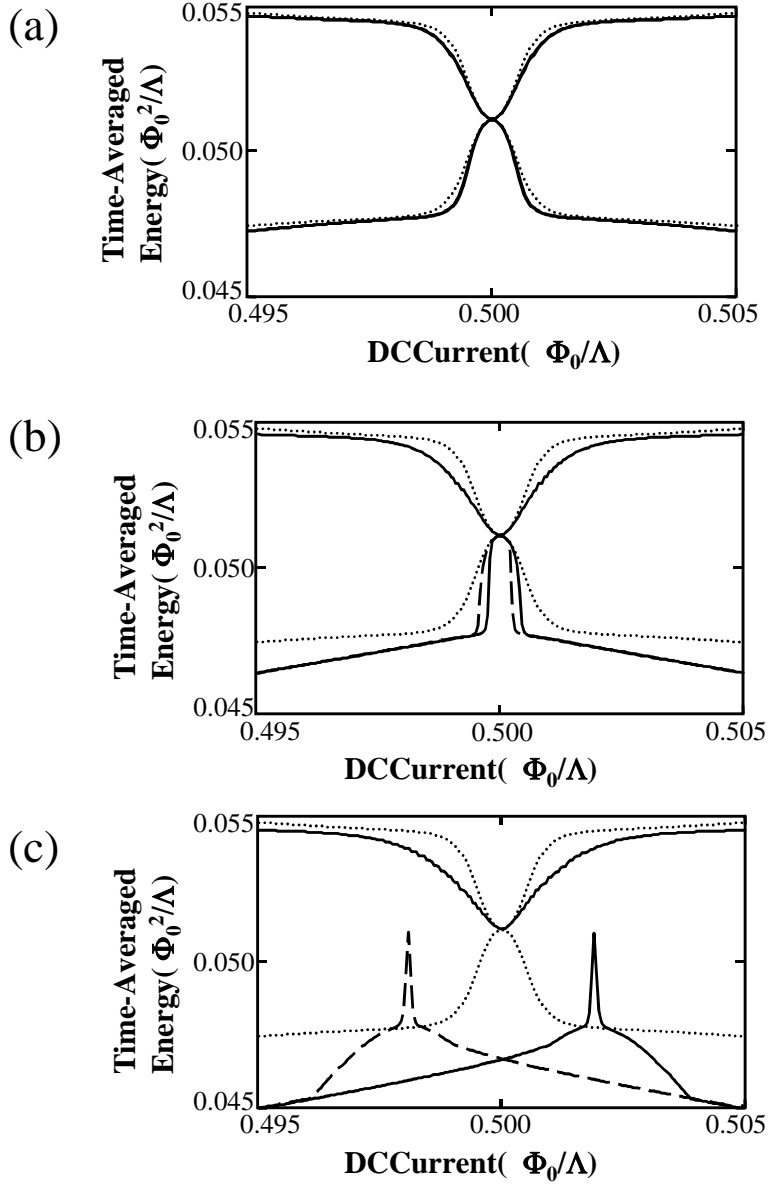}
\end{center}
\caption{Time-averaged energy levels as a function of applied DC current for a DC coil with $L_{dc}=1\times 10^{-9}$H:
(a) $K^2=0.01$ and $\mu=0.055$, (b) $K^2=0.03$ and $\mu=0.095$, (c) $K^2=0.05$ and $\mu=0.122$. The 
original energies shown as a dotted line for comparison. In (b) and (c) increasing current is shown as 
a solid line and decreasing current is shown as a dashed line.}
\end{figure}

The three examples given in figure 2 show very different behavior. The first, figure 2(a) with $K^2=0.01$ corresponding
to $\mu=0.055$, shows that the width of the resonance in the ground state is reduced by the effect of the 
backreaction, whilst the first excited state is broadened. Of course, it would normally be much easier 
to see the ground state behavior than the excited states because of environmental effects, but we
include both for completeness. In figure 2(b), $K^2=0.03$ and $\mu=0.095$, a hysteresis loop appears on either
side of the resonance. This is because there are multiple solutions to equation (\ref{dc}), leading to multiple allowable flux values 
for a single value of the DC current. The situation is even more extreme in figure 2(c), $K^2=0.05$ and $\mu=0.122$, where
the nonlinearities are now so severe that the ground state resonance no longer looks anything like the original. The single
resonance shown in the inset of figure 1 has been split, giving asymmetric resonances that depend on whether
the DC current is being ramped (quasi-statically) up or down. In each of these figures, the first excited state behavior
is relatively consistent. This is because the effect of the backreaction is to broaden the features, reducing the effect
of the nonlinearity rather than narrowing the resonance, which increase the apparent size of the screening current
(which is related to the derivative of the energy level via equation (\ref{sc})). It should be stressed that the behavior
shown in figure 2 is not dependent on the particular solution of the time-dependent Schr\"{o}dinger equation chosen for this 
paper. The same type of behavior should be seen in the region of any quantum transition or resonance with the same
general shape as that shown in the inset of figure 1. Equally, if the time-dependent field is removed the effect of the
nonlinear backreaction will be very much reduced because it is dependent on the curvature of the time-averaged energy
in the SQUID ring. When the time-dependent field is applied, the effect of the nonlinearity is much stronger
because of the 'sharp' shape of the resonance compared to the background curvature of the time-independent 
energy levels (see figure 1).

Although this is an interesting nonlinear effect in its own right, the importance of the backreaction is mainly in 
what it tells us about the ability to control or bias a quantum mechanical SQUID at or near to one of its transitions.
Near to a transition, the (time-averaged) screening currents generated by the ring are very nonlinear and can 
generate very strong nonlinear behavior, such as hysteresis. If one were to try to hold the system near to 
one of these regions of strongly nonlinear behavior, the DC flux might not behave in a predictable manner when
subject to small amounts of noise, with the system `hopping' around between the different possible flux states.
This could cause problems if the system was required to operate in one of these regimes to create quantum entanglements
between elements in a quantum circuit. The only way to reduce these effects is to reduce the coupling between 
the DC coil and the SQUID ring. (The behavior shown in figure 2 is only weakly dependent 
on the individual inductances for the SQUID and the DC coil). Although it may be possible to reduce the couplings 
for individual SQUID systems, there may be practical limitations to this approach when 
designing large scale systems with many SQUID devices, of the type required for a large scale
quantum computer based on SQUID technology. If the coupling is reduced between a SQUID and its DC coil,
to reduce the effects of the backreaction, it becomes difficult to decouple the DC coil from the neighboring 
SQUIDs, which may introduce problems with cross coupling between qubits and/or additional unwanted 
environmental effects. 

\section{Conclusions}

The subject of this paper has been the control of quantum mechanical SQUID rings at, or near to, a transition using an
external static magnetic flux. We have shown that the effect of the backreaction of the SQUID on the DC coil 
can be significant near to a transition, even when the coupling between the two systems is weak. For the example
used in this paper the effect is significant even when the coupling is around 1\% ($K^2=0.01$). The appearance of nonlinear
behavior in the DC coil, such as multiple stable solutions and hysteresis, could lead to unpredictable behavior
in the SQUID and disturb the correct operation of a qubit/quantum gate based on SQUID devices.

\vspace{1cm}
The authors would like to thank the United Kingdom National Endowment for Science Technology and the Arts (NESTA) 
and the Engineering and Physical Science Research Council (EPSRC) Quantum Circuits Network for their generous support.


\end{document}